\documentclass[aps,prb,reprint, groupedaddress]{revtex4-1}
\usepackage{graphicx}

\begin{document}


\title{Direct observation of metastable hot trions in an individual quantum dot}


\author{V. Jovanov}
\email[]{jovanov@wsi.tum.de}
\author{S. Kapfinger}
\author{G. Abstreiter}
\author{J. J. Finley}
\affiliation{Walter Schottky Institut, Technische Universit\"{a}t M\"{u}nchen, Am Coulombwall 4, 85748 Garching, Germany}


\date{\today}

\begin{abstract}
Magneto photoluminescence and excitation spectroscopy are used to probe the excited state spectrum of negatively charged trions in a InGaAs quantum dot. A single dot optical charging device allows us to selectively prepare specific few ($1e$, $2e$) electron states and stabilize hot trions against decay via electron tunneling from excited orbital states. The spin structure of the excited state results in the formation of metastable trions with strong optical activity that are directly observed in luminescence. Excitation spectroscopy is employed to map the excited singlet and triplet states of the two electron wavefunction and fine structure splittings are measured for the lowest lying and excited orbital states. Magneto-optical measurements allow us to compare the g-factors and diamagnetic response of different trion states.
\end{abstract}


\maketitle

%


Semiconductor quantum dots (QDs) have attracted much attention over the past decade due to their potential for realizing new generations of coherent quantum devices.\cite{Hanson2007}  In particular, the spin of charges trapped in QDs exhibits slow relaxation and robust coherence that renders it highly promising for realizing quantum bits in the solid state.\cite{Heiss2010,Gerardot2008}  Moreover, spin can be optically initialized, manipulated and read-out via the singly charged exciton (trion) transitions of the dot.  Whilst several spectroscopic studies have been performed on the lowest energy trions\cite{Warburton2000,Finley2001,Findeis2001,Bayer2002} comparatively few reports have appeared pertaining to \textit{excited} states, so-called \textit{hot} trions.\cite{Ware2005,Warming2009,Sanada2009} Slow spin relaxation renders triplet hot trion states metastable and non-radiative processes such as tunneling from excited orbital states may occur over timescales faster than the radiative recombination, preventing their optical investigation.
Besides the application to optical spin-manipulation and readout\cite{Heiss2010} the fine structure of hot-trions provides rich information about $e-e$ and $e-h$ exchange interactions. The splitting between bright and dark excitons arising from $e-h$ exchange interaction can be measured by breaking the symmetry of the QD with transverse magnetic fields and enhancing the dark exciton oscillator strength.\cite{Bayer2002} Similar information can be obtained directly from the hot trion in combination with the doubly charged exciton without the need of magnetic fields.\cite{Warming2009} In this respect the hot-trions allow fundamental investigations of interactions between (in)distinguishable quantum particles in a solid-state environment.

In this paper we investigate the excited states of the negatively charged trion $X^{-1}$ in individual InGaAs self-assembled QDs.  Using charge storage devices that inhibit electron tunneling we directly observe photoluminescence (PL) from hot trions, contrasting with previous reports where the hot trion was observed \textit{somewhat} indirectly as the final state of the decay of the charged biexciton.\cite{Warming2009,Urbaszek2003} This allows us to perform PL-excitation (PLE) spectroscopy where luminescence from both ground and excited trion states can be simultaneously observed. We find distinct sets of resonances for both states which are identified as excited singlet and triplet states of the two electron wavefunction. From the fine structure splittings of spin-triplet transitions we extract values of the isotropic electron-hole exchange interaction for states in which electrons populate the lowest lying and excited orbital states. Furthermore, magneto-optical measurements allow us to compare the g-factors and diamagnetic response of different trion quantum states.

\begin{figure}[htb]
\includegraphics[width=\columnwidth]{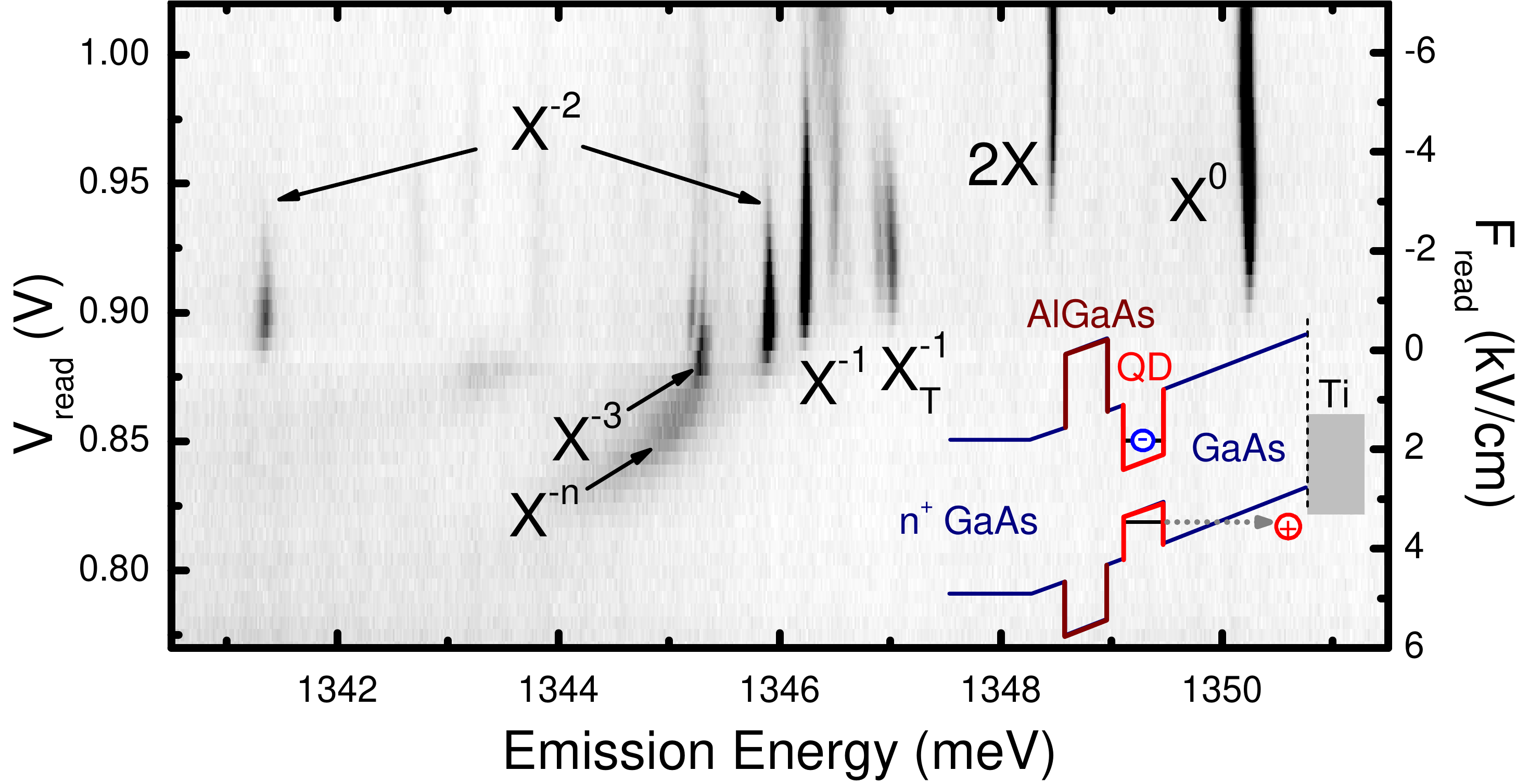}
\caption{\label{Figure01} Typical image plot of PL spectra obtained for various voltage biases applied in growth direction.}
\end{figure}

%
The samples investigated incorporated a low density ($\leq5$~$\mu$m$^{-1}$) layer of nominally In$_{0.5}$Ga$_{0.5}$As/GaAs self-assembled QDs into the $140$~nm thick intrinsic region of \emph{n}-\emph{i}-Schottky photodiode structures, $40$~nm above the $n^{+}$ ($1\times10^{18}$~cm$^{-3}$) GaAs contact layer.\cite{Heiss2008} An asymmetric Al$_{0.45}$Ga$_{0.55}$As tunnel barrier with thickness of $20$~nm was grown immediately below the QDs, allowing for the optical generation and storage of electrons inside the dots.\cite{Heiss2009,Heiss2010} Individual QDs were spatially isolated via $1$~$\mu$m diameter shadow mask apertures made in a semitransparent Ti-Au top metallic contact. The corresponding band diagram of the device is schematically depicted in the inset of fig.~\ref{Figure01}.

Single dots were optically probed using a low temperature ($10$~K) magneto-confocal microscopy set up that allows magnetic ($B$) fields up to $B=15$~T to be applied in Faraday geometry. Optical excitation was achieved using two independently tunable Ti:Sapphire lasers that were switched on and off using acousto-optic modulators. The optical excitation pulse sequence was temporally synchronized with the electric field applied to the device and the devices had a switching bandwidth in excess of $1$ MHz. The emitted PL-signal from single QDs was analyzed using a double ($2\times0.5$~m) spectrometer and detected by a LN$_{2}$ cooled Si CCD camera.

Typical PL spectra recorded from an individual QD are presented in the image plot in fig.~\ref{Figure01} as a function of the voltage $V_\mathrm{read}$ applied to the device. Using the expected built-in potential of the Schottky junction, $V_\mathrm{read}$ can be converted to the static electric field ($F_\mathrm{read}$) experienced by the dots. The sign convention adopted in this paper is such that $F_\mathrm{read}$ orientated anti-parallel (parallel) to the QD growth direction is negative (positive). The data presented in fig.~\ref{Figure01} were recorded with non-resonant excitation ($1437$~meV) into the wetting layer (WL) continuum. Due to the presence of the AlGaAs tunnel barrier, \textit{electrons} tend to accumulate in the dot and WL since photogenerated holes are removed over timescales comparable to the radiative lifetime (inset in fig.~\ref{Figure01}), whilst escape of electrons is blocked by the AlGaAs barrier.\cite{Heiss2008}
As the electric field shifts from negative to positive polarities, the balance between electron trapping, hole removal and depletion by radiative recombination changes. For positive $F_\mathrm{read}$ the dots contain photogenerated electrons and emission from highly negative excitons $X^{-n}$ (fig.~\ref{Figure01}) dominates the PL spectra. In contrast, as $F_\mathrm{read}$ is shifted from positive to negative polarity, geminate $e-h$ capture into the dot becomes increasingly favored, reducing the electron occupancy and resulting in the appearance of characteristic triply ($X^{-3}$), doubly ($X^{-2}$) and singly ($X^{-1}$) charged excitons in the time integrated spectrum. Neutral excitonic emission lines ($X^{0}$, $2X$) emerge as $F_\mathrm{read}$ becomes more negative than $-1$~kV/cm. In this situation the AlGaAs layer increasingly presents a tunnel barrier for holes, increasing the tunneling times such that the electron-hole pair capture into the dot becomes predominantly geminative. We assign the doublet emission in the PL spectra labeled as $X^{-1}_{T}$ to the decay of the triplet states of the hot negatively charged trion, an assignment that is firmly substantiated below. Of course, whilst Pauli exclusion requires that an electron populates a higher orbital state for $X^{-1}_{T}$, the complex is stabilized by the AlGaAs barrier in the samples studied here.

%
\begin{figure}[tb]
\includegraphics[width=\columnwidth]{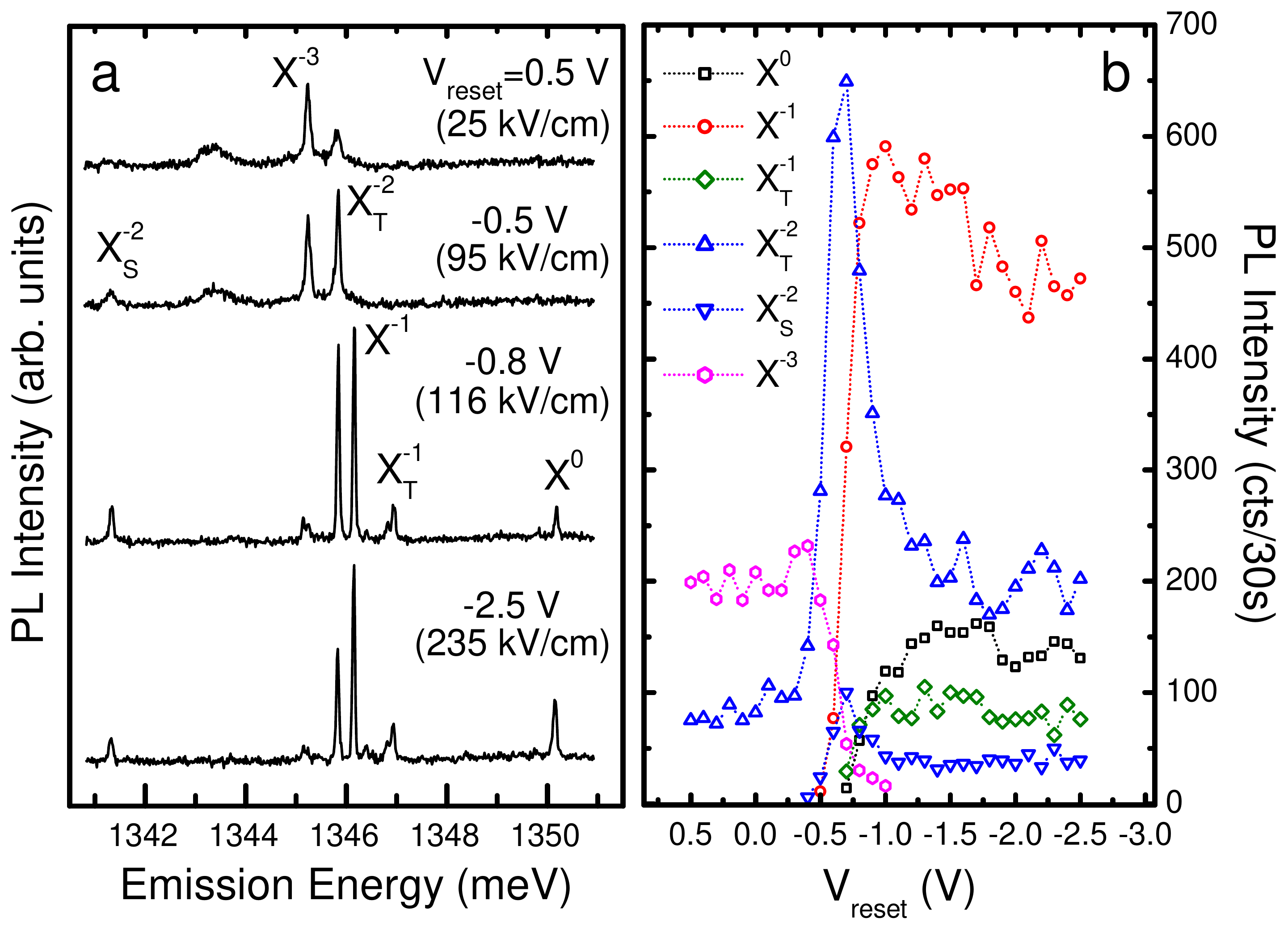}
\caption{\label{Figure02} (Color online) (a) Evolution of the PL spectrum for fixed readout voltage $V_\mathrm{read}=0.86$~V ($F_\mathrm{read}=0$~kV/cm) and reducing (increasing) reset voltage (electric field) pulse with a duration of $500$~ns. (b) PL intensity of the individual excitons from (a) as a function of the reset voltage.}
\end{figure}

To firmly substantiate the identification of the transitions in fig.~\ref{Figure01} we performed control experiments where the QD was periodically emptied of \textit{all }charge prior to the application of the readout electric field $F_\mathrm{read}$ and optical measurement. This was achieved by application of a strong voltage pulse $V_\mathrm{reset}$ that generates an electric field pulse $F_\mathrm{reset}$ during which accumulated electrons tunnel out from the dot, despite the presence of the tunnel barrier.\cite{Heiss2008} The influence of the reset voltage pulse on the PL spectra is illustrated in fig.~\ref{Figure02}(a) and the detailed systematic dependence of the PL intensity of the individual transitions is plotted in fig.~\ref{Figure02}(b) as a function of $V_\mathrm{reset}$. In these measurements, the readout voltage was maintained at $0.86$~V ($F_\mathrm{read}=0$~kV/cm) and the $500$~ns duration reset voltage pulse $V_\mathrm{reset}$ was applied periodically ($f=400$~kHz) with an amplitude that was systematically varied from $0.5$~V to $-2.5$~V ($25 - 235$~kV/cm).
The readout laser was switched off during the \textit{reset} phase of the measurements and turned on during the \textit{readout} for $1$~$\mu$s while the PL signal was recorded.
For $V_\mathrm{reset}=0.5$~V ($25$~kV/cm), when the electron tunneling times from the dot are much longer than the duration of the reset pulse, the dot is negatively charged and the PL spectrum is dominated by the triply negatively charged exciton $X^{-3}$ (fig.~\ref{Figure02}(a) and (b)). As $V_\mathrm{reset}$ reduces to $-0.5$~V ($F_\mathrm{reset}$ increases to $95$~kV/cm) the electrons tunnel out of the dot leading to reduction of the average electron occupation, manifesting itself as a decrease of the intensity of $X^{-3}$. At the same time the intensity of the very characteristic\cite{Warburton2000} singlet and triplet transitions of the doubly charged exciton ($X^{-2}_{S}$ and $X^{-2}_{T}$) increases. Continuing this trend, the singly negatively charged excitons $X^{-1}$ and $X^{-1}_{T}$ and the neutral exciton $X^{0}$ appear in the PL spectra as $V_\mathrm{reset}$ reduces below $-0.57$~V ($F_\mathrm{reset}$ increases above $100$~kV/cm) and the average electron occupation in the dot reduces. For $V_\mathrm{reset}<-1.3$~V ($F_\mathrm{reset}>150$~kV/cm) the dot is completely emptied of electrons during the reset phase of the measurement but electrons can still be generated during the readout, giving rise to emission from negative trions. 
We note that the PL intensity of $X^{-1}_{T}$ and $X^{-1}$ have the same characteristic dependence on $V_\mathrm{reset}$ clearly establishing the doublet $X^{-1}_{T}$ to the decay of a singly negatively charged exciton. We have also performed power dependent PL measurements (not presented) on $X^{-1}_{T}$ which revealed a clear linear dependence of the PL intensity on the excitation power, confirming that it does not arise from a multi-exciton transition.

%
\begin{figure}[tb]
\includegraphics[width=\columnwidth]{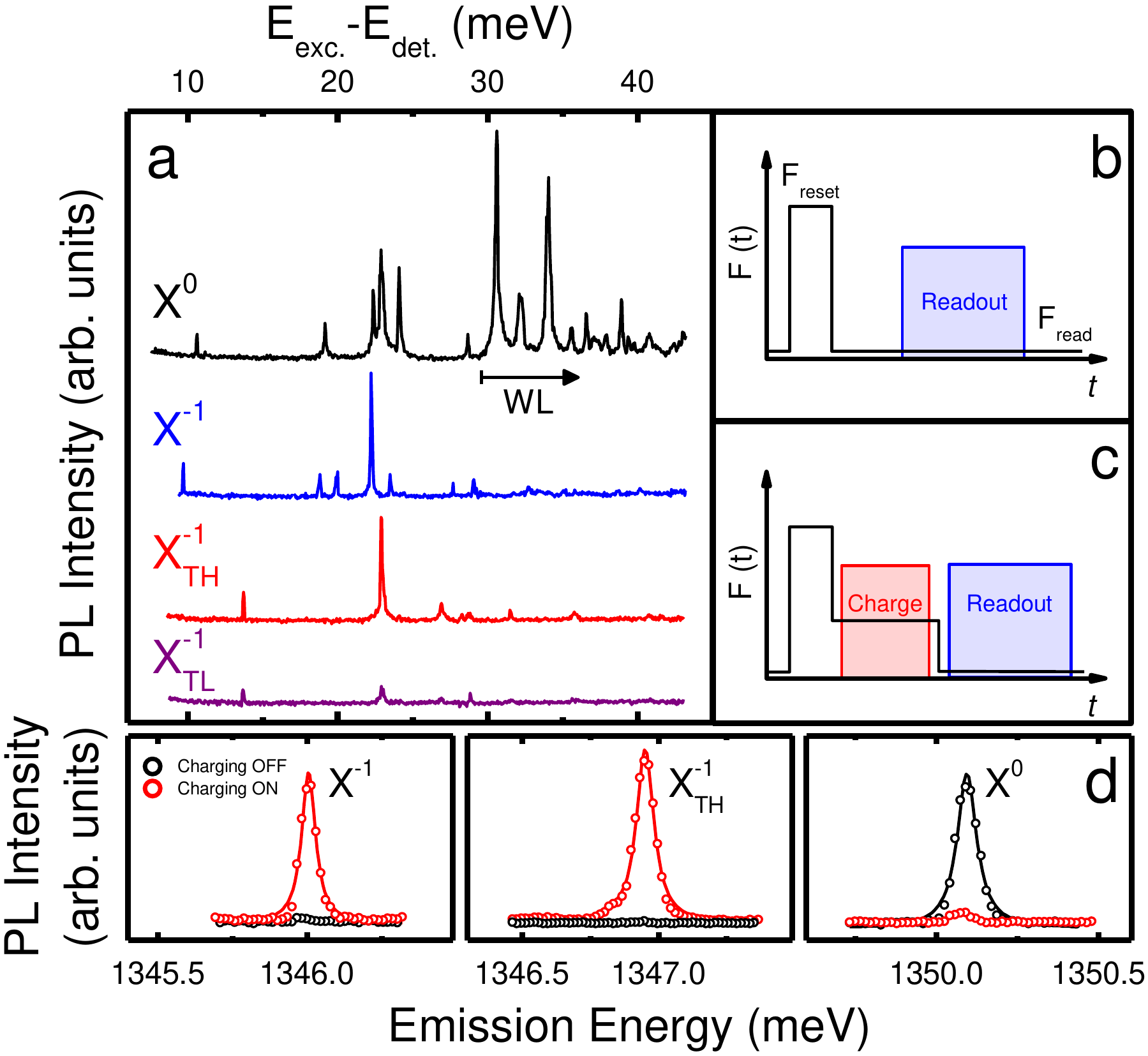}
\caption{\label{Figure03} (Color online) (a) Photoluminescence excitation spectra of the neutral exciton $X^{0}$, the singly charged exciton $X^{-1}$ and the two emission lines of $X^{-1}_{T}$. The emission at higher (lower) energy of $X^{-1}_{T}$ is labeled as $X^{-1}_{TH}$ ($X^{-1}_{TL}$). The electric field sequence applied on the device while obtaining the $X^{0}$ spectrum is shown in (b). The corresponding electric field sequence for $X^{-1}$, $X^{-1}_{TH}$ and $X^{-1}_{TL}$ is shown in (c). (d) PL spectra of $X^{-1}$, $X^{-1}_{TH}$ and $X^{0}$ under resonant excitation in higher orbital states with the electric field sequence (c) applied.}
\end{figure}

In order to charge the dot with a precise number of electrons in a controlled manner, the charge occupancy must not be disturbed during the readout phase of the measurement. This is achieved by resonant excitation of discrete excited orbital states for which the subsequent carrier capture efficiency into the dot is unity. We identified suitable excited orbital states of the studied QD by performing PL-excitation (PLE) spectroscopy. Typical PLE spectra recorded from the neutral exciton $X^{0}$ and the singly negatively charged excitons ($X^{-1}$ and $X^{-1}_{T}$) are compared in fig.~\ref{Figure03}(a). As already seen from the PL measurements presented in fig.~\ref{Figure02}(a), $X^{-1}_{T}$ is actually a doublet and the higher energy line is labeled as $X^{-1}_{TH}$ and the lower energy transition is labeled as $X^{-1}_{TL}$ in fig.~\ref{Figure03}(a), respectively. The electric field sequence and the laser pulse applied to the device whilst recording the $X^{0}$ spectrum are depicted schematically in fig.~\ref{Figure03}(b). The readout ($F_\mathrm{read}$) and reset ($F_\mathrm{reset}$) electric fields were kept at $4.2$~kV/cm and $200$~kV/cm, respectively. In the PLE spectrum recorded from $X^{0}$ (fig.~\ref{Figure03}(a)) several groups of discrete resonances can be identified. The lowest states appear $\approx 10$~meV above the ground state transition $E(X^{0})=1348.78$~meV, while other higher orbital states of the QD are detected up to $\approx 30$~meV above the detected state. For larger excitation energies, a complex series of resonances appear that most likely reflect weakly localized neutral exciton states in the underlying WL.\cite{Brunner1994} This suggestion is supported by the spectral position and width of the WL PL emission that is centered at $1396$~meV.

For recording the PLE spectra of the singly negatively charged excitons (fig.~\ref{Figure03}(a)) the QD was optically \textit{pre}charged with an additional laser before the PLE measurement was made.  To do this, a $500$~ns duration charging pulse was tuned into resonance with the higher orbital states of $X^{0}$ at electric fields of the order of $100$~kV/cm where a high charging efficiency is ensured. The scheme of the applied electric field sequence and laser pulses is shown in fig.~\ref{Figure03}(c). During the measurement cycle $F_\mathrm{read}$ and $F_\mathrm{reset}$ were identical to the $X^{0}$ measurement. The PLE spectrum of $X^{-1}$ (fig.~\ref{Figure03}(a)) consists of the same number of resonances as $X^{0}$, but red-shifted in energy. This red-shift arises from the renormalization of the excitonic transition energies due to the Coulomb interaction.\cite{Warburton2000} In the case of $X^{-1}$ the attraction between the two electrons and the hole outweighs the electron-electron repulsion by an amount larger than the attraction between one electron and one hole. On the other hand, the PLE spectra of $X^{-1}_{TH}$ and $X^{-1}_{TL}$ reveal a smaller number of correlated resonances. Additional proof that the $X^{-1}_{T}$ is a singly charged negative exciton is obtained from its response to the charging laser. The effect of the presence/absence of the charging laser on the PL from $X^{-1}$, $X^{-1}_{TH}$ and $X^{0}$ is presented in fig.~\ref{Figure03}(d). When the readout laser is tuned into resonance with one of the higher orbital states of $X^{-1}$ and $X^{-1}_{TH}$, PL is generated only if the charging laser is switched \textit{on} prior to the readout, i.e.\ when the dot is precharged with one electron. The opposite is found for the $X^{0}$ PL which is completely quenched when the dot is precharged and restored when the charging laser is \textit{off} and the dot remains charge neutral. We also note that the complex series of resonances observed in the spectral vicinity of the WL for $X^{0}$ are not observed for the negatively charged states. This observation supports our identification of the negatively charged excitons, since a pre charged dot is more likely to non-geminately capture an additional hole, producing luminescence from the $X^{0}$ transition. 

%
\begin{figure}[tb]
\includegraphics[width=\columnwidth]{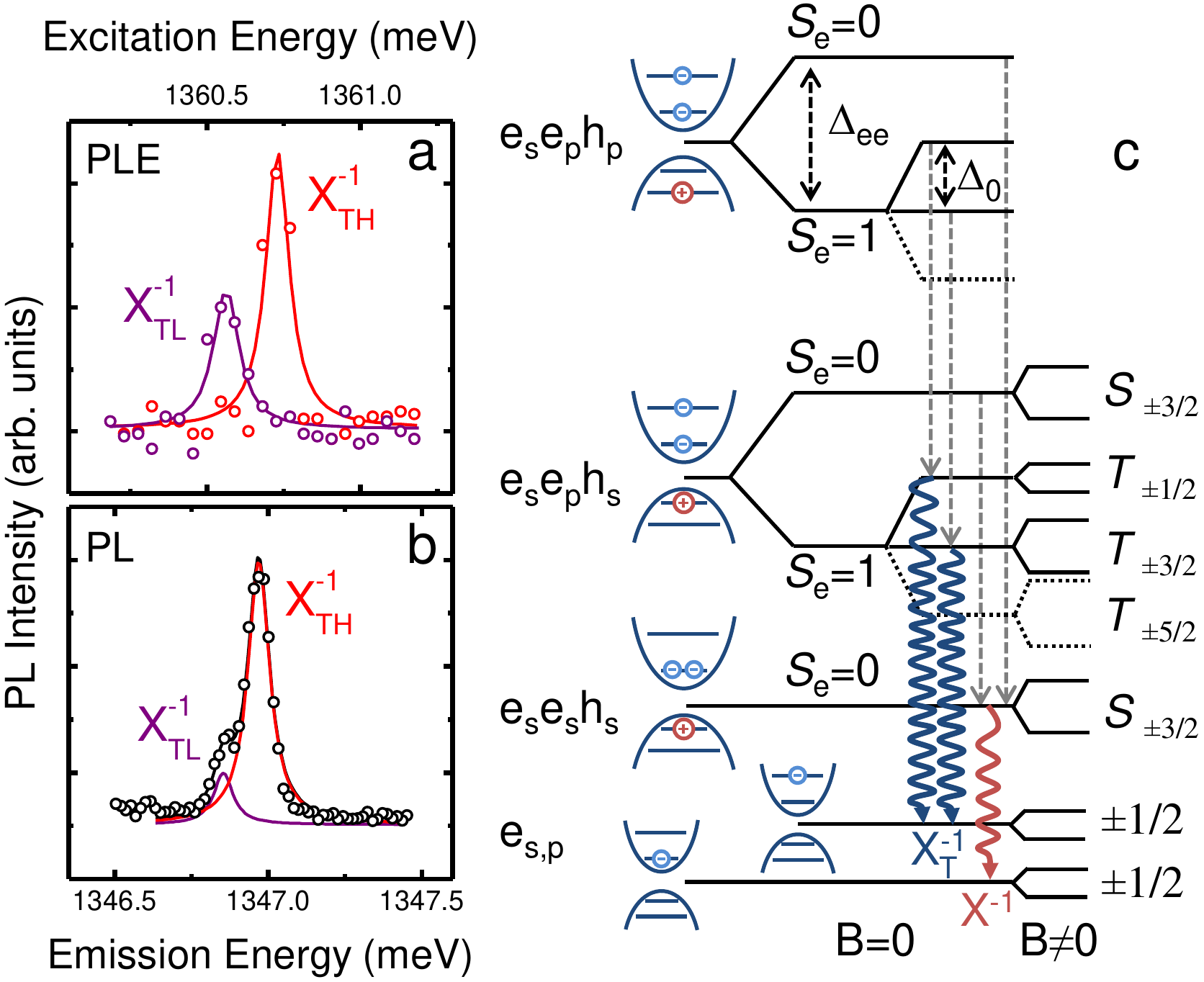}
\caption{\label{Figure04} (Color online) (a) Photoluminescence excitation spectra of the energetically lowest resonances of $X^{-1}_{TH}$ and $X^{-1}_{TL}$ and (b) their photoluminescence spectrum under WL excitation. (c) Energy level diagram of two electrons and one hole in a QD. The optical transitions are indicated by the wavy arrows, the non-radiative transitions with straight broken lines.}
\end{figure}


We continue by focusing on the PLE spectra recorded from $X^{-1}_{TH}$ and $X^{-1}_{TL}$ shown in fig.~\ref{Figure03}(a). The spectra reveal two striking features: The number of resonances is much smaller compared to $X^{0}$ and $X^{-1}$, and $X^{-1}_{TH}$ and $X^{-1}_{TL}$ each have same number of resonances shifted in energy ($90 - 170$~$\mu$eV) with respect to each other, as for their corresponding emission lines. This splitting can be seen more clearly in fig.~\ref{Figure04}(a) that shows the PLE spectra of the energetically lowest resonances of $X^{-1}_{TH}$ and $X^{-1}_{TL}$. In the same way there is a splitting in their PL emission of $115\pm7$~$\mu$eV, shown in fig.~\ref{Figure04}(b). In order to understand the fine structure of the trion excited states, we consider the expected level structure as shown in fig.~\ref{Figure04}(c). We attribute the observed splitting between the PL and PLE resonances of $X^{-1}_{T}$ as arising from the isotropic electron-hole exchange interaction $\Delta_0$ which lifts the degeneracy of the triplet states (total electron spin $S_{e}=1$),\cite{Akimov2002,Kavokin2003} as depicted schematically in fig.~\ref{Figure04}(c). The difference of the splittings in emission and excitation originates from the different orbital states occupied by the electrons and the hole. In excitation one electron resides in the lowest orbital state, while the additional electron-hole pair is generated in higher orbital state. In contrast, for emission the recombination always occurs between an electron and a hole occupying s-orbitals, while the additional electron resides in the p-orbital.
In addition, the excited singlet levels ($S_{e}=0$) are split from the corresponding triplet states by the electron-electron exchange interaction $\Delta_{ee}$.\cite{Akimov2002}
Using the energy level scheme illustrated in fig.~\ref{Figure04}(c) it follows that the PL emission labeled as $X^{-1}_{TH}$ originates from the radiative decay of the $X^{-1}_{T \pm1/2}$ state with total spin $\pm1/2$ and, analogously, $X^{-1}_{TL}$ arises from the radiative decay of $X^{-1}_{T \pm3/2}$ with total spin of $\pm3/2$. The triplet state $X^{-1}_{T \pm5/2}$ is optically inactive due to spin conservation  and is not observed as expected.\cite{Bayer2002} The splitting of the two triplet states $\Delta_{0}$ is a sum of the electron-hole exchange interactions between the s-hole and the  s- and p- electrons $2\Delta_{0}=(\Delta_{0}^{e_{s}h_{s}}+\Delta_{0}^{e_{p}h_{s}})$.\cite{Urbaszek2003} Here, we neglect the splittings due to anisotropic exchange interactions, since for the studied QD they are one order of magnitude smaller than the isotropic interactions \cite{Klotz2010} and are unmeasurable due to the limited experimental resolution ($\approx40$~$\mu$eV). The electron-hole exchange interaction $\Delta_{0}^{e_{p}h_{s}}$ can be readily obtained from the triplet state of the doubly charged exciton $X^{-2}_{T}$, since in the initial state it has two spin paired s-electrons and an unpaired p-electron and s-hole. Since we do not observe a splitting of $X^{-2}_{T}$ (fig.~\ref{Figure01} and fig.~\ref{Figure02}(a)) it follows that the electron-hole exchange interaction between a p-electron and s-hole is much smaller than the experimental resolution $\Delta_{0}^{e_{p}h_{s}}\ll40$~$\mu$eV. Thus, we calculate the electron-hole exchange interaction between the s-electron and s-hole, obtaining $\Delta_{0}^{e_{s}h_{s}}\approx230$~$\mu$eV in excellent accord with previous reports on similar QDs obtained using different methods.\cite{Bayer2002}

During the PLE measurements of $X^{-1}$ radiative recombination occurs only from the energetically lowest singlet states via emission of $X^{-1}$ photons since electrons in excited singlet configurations rapidly relax into the lowest singlet level (fig.~\ref{Figure04}(c)). Hence, the  resonances in the PLE spectrum of $X^{-1}$ (fig.~\ref{Figure03}(a)) represent the excited singlet levels of two electrons with anti parallel spins. Analogously to the excited singlets, the electrons and the hole from the excited triplets relax to the energetically lowest triplet state from where radiative recombination occurs. As mentioned above, the excited singlet and triplet states are split by the electron-electron exchange interaction. The same interaction produces the splitting of $4.5$~meV between the singlet and the triplet final states of the doubly charged negative exciton $X^{-2}$ (fig.~\ref{Figure01} and fig.~\ref{Figure02}(a)).\cite{Warburton2000} A close inspection of the PLE spectra of the negatively charged excitons shown in fig.~\ref{Figure03}(a) indeed reveals that $4.3$~meV above the energetically lowest resonance of $X^{-1}_{T}$ there is also a resonance in the PLE spectrum of $X^{-1}$, as expected. The same applies for the second excited state resonance of $X^{-1}_{T}$ above which $X^{-1}$ resonance is present $4.0$~meV to higher energy. When considered individually and together, these observations strongly substantiate our identification of $X^{-1}_{T}$ as arising from the hot-trion with triplet spin character.


%
\begin{figure}[tb]
\includegraphics[width=\columnwidth]{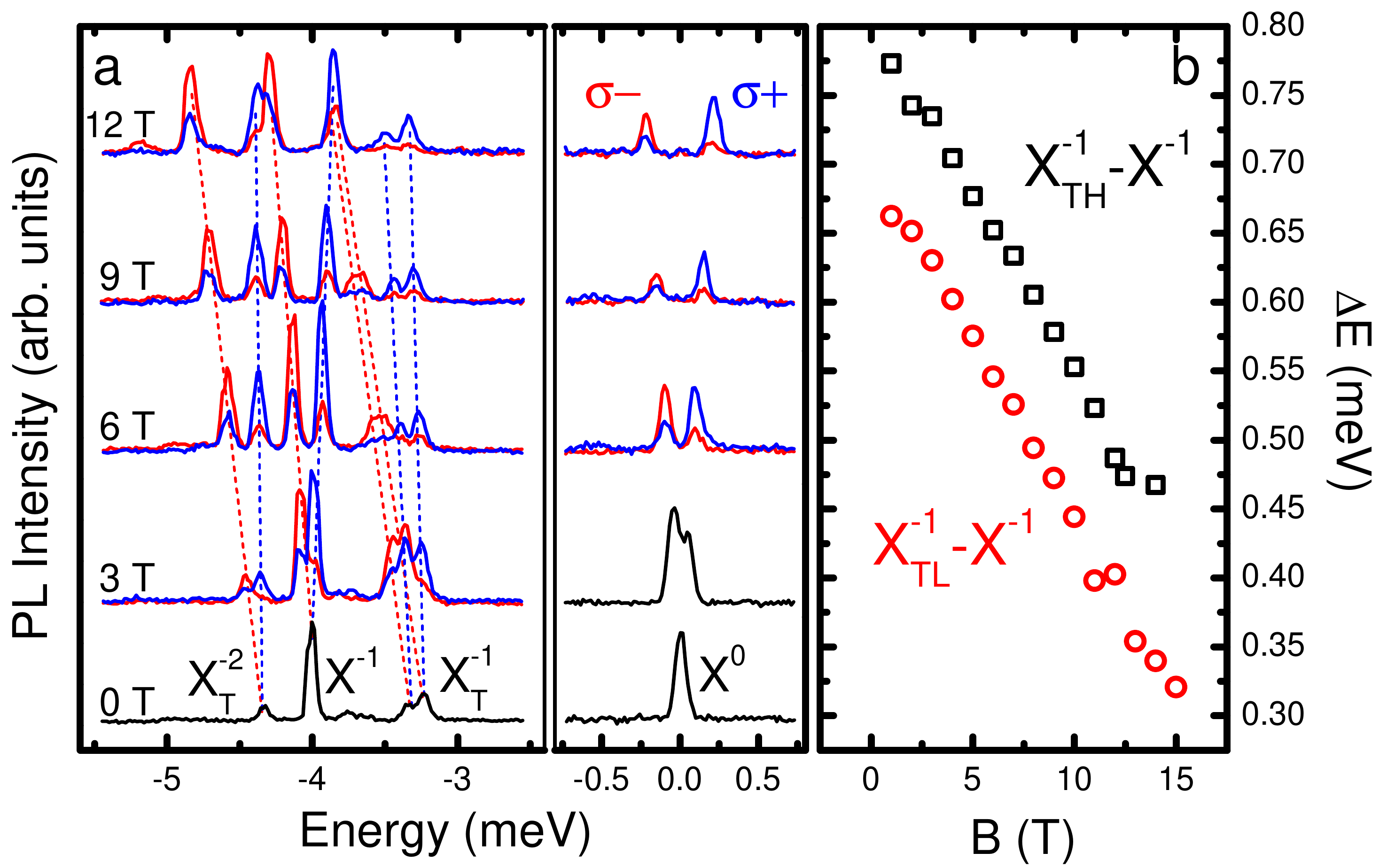}
\caption{\label{Figure05} (Color online) (a) Polarization resolved PL spectra of $X^{-2}_{T}$, $X^{-1}$, the two lines of $X^{-1}_{T}$, and $X^{0}$ as a function of magnetic field applied in growth direction. (b) Magnetic field dependence of the energy gap between the $X^{-1}_{T}$ and $X^{-1}$ transitions.}
\end{figure}


To further substantiate our peak assignments we investigated the magneto-optical response of the hot trion.
Fig.~\ref{Figure05}(a) compares PL spectra from the neutral exciton $X^{0}$, the singlet state of the negative trion $X^{-1}$, the two trion triplet states $X^{-1}_{T}$ and the triplet state of the doubly charged exciton $X^{-2}_{T}$ as a function of magnetic field applied in Faraday configuration.
For clarity, the spectra are translated by the diamagnetic shift of the neutral exciton $X^{0}$. As the magnetic field is increased, each of the PL lines splits into two circularly polarized transitions, due to Zeeman splitting of the electron and the hole. For all of the transitions, we extracted excitonic g-factors of the order of $g_{ex}=0.60\pm0.01$. Since all of the studied excitons have either one electron  or one hole either in the initial or in final state, their excitonic splittings are the same as expected within the peak identification framework presented in this paper. A marked exception is the splitting of the $X^{-1}_{T\pm3/2}$ ($X^{-1}_{TL}$) where the electron part of the splitting arises from the electron in the p-state. Since we measured the same splitting for this exciton as for the other transitions studied we conclude that either the s- and p-electron g-factors are very similar or the difference is compensated by magnetic field dependent Coulomb interactions.\cite{Schulhauser2002}

From Fig.~\ref{Figure05}(a) it is evident that the diamagnetic response of the triplet exciton emissions ($X^{-1}_{TH}$ and $X^{-1}_{TL}$) is very different from that observed for $X^{0}$ and $X^{-1}$. $X^{0}$ and $X^{-1}$ exhibited very similar diamagnetic shifts with diamagnetic coefficients $16.5\pm0.1$~$\mu$eV/T$^2$ and $16.2\pm0.1$~$\mu$eV/T$^2$, respectively. Since the additional electron in the dot for the case of $X^{-1}$ does \textit{not} introduce paramagnetic contributions it follows that the excitons in the studied QD are in the strong confinement regime and the single particle energies are much larger than the Coulomb interactions.\cite{Schulhauser2002} On the other hand, the triplet states $X^{-1}_{TH}$ and $X^{-1}_{TL}$ exhibit smaller diamagnetic shifts with similar constants of $14.5\pm0.05$~$\mu$eV/T$^2$ and $14.7\pm0.05$~$\mu$eV/T$^2$. Here, the paramagnetic contribution is in the order of $-2$~$\mu$eV/T$^2$ and is introduced by the additional electron in the triplet states. This can be clearly seen in fig.~\ref{Figure05}(b) where the evolution of the energy gaps between the $X^{-1}_{TH}$, $X^{-1}_{TL}$ and $X^{-1}$ transitions are presented upon increasing the magnetic field. Due to the existence of exchange density in the triplet states, electrons tend to spatially avoid each other\cite{Sakurai1994} leading to a reduction of the Coulomb repulsion, a direct consequence of the requirement of an antisymmetrical wave function for the electron. In the case of the energetically lowest singlet trion $X^{-1}$ the Coulomb repulsion is stronger since both electrons reside closer in the s-orbital state. An increasing magnetic field acts as additional confinement which reduces the lateral extension of the two electron wave function. This reduces the separation between the electrons and in the case of the lowest singlet state, when the electrons are closer to each other, the magnetic field dependent Coulomb repulsion grows stronger than in the case of the triplet states where one electron is in the energetically higher p-orbital state. This effect has been theoretically predicted \cite{Wagner1992} and experimentally observed in transport measurements on gated few-electron QDs.\cite{Kouwenhoven1997}

%
%
In summary, we have used QD electron storage devices to directly study metastable states of the negative trions. By performing photoluminescence and excitation spectroscopy we probed the excitation spectrum of singlet and triplet states of the two electron wavefunction and measured the fine structure splittings. From magneto-optical measurements we obtained the g-factors and diamagnetic response of the different trion states.\\
\begin{acknowledgments}
This work is funded by the DFG via SFB-631 and the excellence cluster Nanosystems Initiative Munich (NIM) and the European Union via SOLID. G. A. also thanks the TUM Institute for Advanced Study for support.
\end{acknowledgments}

\bibliography{Neg-Trion_bib}

\end{document}